\documentclass{article}
\usepackage{amssymb}
\usepackage{graphicx}
\usepackage{amsfonts}
\usepackage{amsmath}
\usepackage{latexsym}
\usepackage{epsfig}
\usepackage{color}
\usepackage{multirow}
\usepackage{mathrsfs}
\begin{document}
 \newcommand{\Tr}{\mathop{\rm Tr}\nolimits}
\newcommand{\para}{_\parallel}
\newcommand{\pr}{_\perp}
\newcommand{\fs}{\rlap/}
\def\twidle{\widetilde}
\def\f{\frac}
\def\omit#1{_{\!\rlap{$\scriptscriptstyle \backslash$}
{\scriptscriptstyle #1}}}
\def\vec#1{\mathchoice
    {\mbox{\boldmath $#1$}}
    {\mbox{\boldmath $#1$}}
    {\mbox{\boldmath $\scriptstyle #1$}}
    {\mbox{\boldmath $\scriptscriptstyle #1$}}
}
\def\eqn#1{Eq.\ (\ref{#1})}
\def\boxit#1{\vcenter{\hrule\hbox{\vrule\kern8pt
      \vbox{\kern8pt#1\kern8pt}\kern8pt\vrule}\hrule}}
\def\Boxed#1{\boxit{\hbox{$\displaystyle{#1}$}}} 
\def\sqr#1#2{{\vcenter{\vbox{\hrule height.#2pt
        \hbox{\vrule width.#2pt height#1pt \kern#1pt
          \vrule width.#2pt}
        \hrule height.#2pt}}}}
\def\square{\mathchoice\sqr34\sqr34\sqr{2.1}3\sqr{1.5}3}
\def\Square{\mathchoice\sqr67\sqr67\sqr{5.1}3\sqr{1.5}3}
\def\lambdabar{{\mathchar'26\mkern-9mu\lambda}}
\def\thrdotovervx{\buildrel\textstyle...\over v_x}
\def\thrdotovervy{\buildrel\textstyle...\over v_y}
\title{\bf Electron Dynamics in Noncommutative Geometry with Magnetic Field and Zitterbewegung Phenomenon}
\author{{\small \ Mehran Zahiri Abyaneh}\footnote{me\_zahiri@sbu.ac.ir} \ {\small and\
         Mehrdad Farhoudi}\footnote{m-farhoudi@sbu.ac.ir}\\
        {\small Department of Physics, Shahid Beheshti University, G.C.,
        Evin, Tehran 19839, Iran}}
\date{\small June 20, 2021}
\maketitle
\begin{abstract}
\noindent
\end{abstract}
Starting from a gauge invariant Dirac Hamiltonian with noncommutativity of space sector in
the presence of an external uniform magnetic field, the resulting Dirac equation has been
solved for electrons and its corresponding \emph{zitterbewegung} ({\bf
ZBW}) phenomenon has been studied. The corresponding energy spectrum is shown to be different from previous studies wherein the non-gauge invariant Dirac Hamiltonian has been used.
The effects of noncommutativity alter the amplitude as well as the cyclotron
and the ZBW frequencies of the average velocity of charge
carriers. This result is contrary to previous studies wherein there was no magnetic field and hence,
neither the amplitude nor the frequency of the motion was affected.
Moreover, all of the ZBW frequencies of the Landau energy-levels
appear in the results. Also, in weak magnetic fields, we have
calculated the average velocity of charge carriers for two
initially localized spin-up and spin-down cases. The
plotted trajectories reveal difference
between these two cases. In addition, the ZBW phenomenon has been
shown that manifests itself as a circular motion whose direction
is spin dependent while accompanied by the cyclotron motion.\\
\medskip
{\small \noindent
 PACS number: 02.40.Gh; 03.65.-w; 11.10.Nx;
                               03.65.Sq; 13.40.Em}\newline
{\small
 Keywords: Noncommutative Geometry; Zitterbewegung; Magnetic Field}
\bigskip
\section{Introduction}
\indent

One of the outcomes of the Dirac equation is a phenomenon, which
has been dubbed ZBW by
Schr\"odinger~\cite{schrodinger30-31,barut-etal81a}. It is a
trembling/quivering motion which is supposed to be endured by a
free Dirac particle while indicates that its velocity and momentum
do~not coincide and rapidly oscillates with the speed $c$ around
its center of mass whereas moves like a relativistic particle with
velocity ${\bf p}/m$~\cite{Rose61}--\cite{Merzbacher98}. The
amplitude of such a motion is predicted to be of the order of the
Compton wavelength, i.e. $\lambdabar_c=\hbar/(m_ec)\simeq
3.86\times {10}^{-13}\, {\rm m}$ for electrons. Through this
picture, it has also been suggested that the spin, and in turn the
magnetic moment, of an electron (as a point charge) is generated
as an intrinsic local motion~\cite{huang1952}--\cite{hestenes2009}
such that the direction of its spin is caused via the motion of
particle around a circle of radius $\lambdabar_c/2$ with frequency
$\omega^{\rm [zbw]}(\simeq 1.6\times {10}^{21}\, {\rm s}^{-1}$ for
electrons). This result leads to the intrinsic magnetic moment of
particle with the correct gyromagnetic $g$ factor. Even in this
regard, by taking the neutrino as a localized wave packet
consisting of positive and negative energies, its chiral ZBW has
been studied~\cite{Stefano}. Wherein, it has been claimed that it
may explain the `missing' solar neutrino experiments. Besides, the
chiral oscillations have been interpreted~\cite{Bernardini} in
terms of the ZBW effect. Moreover, the ZBW of massive and massless
scalar (spin-$0$) bosons and massive (spin-$1$) bosons have been
analyzed in Ref.~\cite{Silenko}. Furthermore, the matter of an
electron in the presence of an external magnetic field has widely
been studied, see, e.g.,
Refs.~\cite{ItzyksonZuber}--\cite{Bhattacharya}, where its ZBW
effect has also been considered by us~\cite{zahirifarhoudi}.

In addition, the ZBW phenomenon has been investigated in many
different contexts, one of which is the Dirac
materials~\cite{Balatsky}. Indeed, many of the relativistic
concepts, like the ZBW, have been transferred from the
relativistic quantum mechanics to Dirac materials~\cite{Balatsky}.
Interestingly enough, the phenomenon of ZBW for electrons has also
been simulated using the trapped ions~\cite{Gerristma10}, and it
has been shown (e.g., Refs.~\cite{Ferrari}--\cite{review} and
other references mentioned in
Refs.~\cite{zahirifarhoudi,Sasabe2014}) to occur in
non-relativistic cases and in two-dimensional Dirac materials
(like graphene and silicon~\cite{review,17}--\cite{Romera2}) as
well. In the latter cases, as the speed of light is replaced by
the Fermi velocity\rlap,\footnote{The original Dirac equation is a
Lorentz covariant, but, in this case, the corresponding result
is~not a Lorentz covariant.}\
 it leads to
lower-frequencies and higher-amplitudes for the ZBW motion that
give more hopes of being detected in this framework. In the case
of such a detection, it would indirectly support the occurrence of
the ZBW in vacuum as well.

Nevertheless, it has been claimed~\cite{Foldy} that the ZBW can be
avoided by a unitary Foldy-Wouthuysen transformation (FW), in
which negative energy components in electron wave functions have
been eliminated. In contrast, one should be reminded that the ZBW
can occur when both positive and negative energy solutions are
considered together. Moreover, the results of such a FW
transformation are in contrast with the Darwin term~\cite{Darwin}
in atomic physics. Also, it has been speculated that an observable
ZBW may only occur in curved spacetimes~\cite{Kobakhidzea}.

Another concept in high-energy physics, which can lead to a new
vision for spacetime structure, is the noncommutative ({\bf NC})
geometry, see, e.g., Refs.~\cite{CuFaZa}--\cite{RSFM}. This topic
has been discussed thoroughly in the literature both on
theoretical and phenomenological aspects, see, e.g.,
Refs.~\cite{Hinchliffe,MF12} and references therein. The concept
of noncommutativity has also strong motivation in the framework of
the string and M-theories, see, e.g.,
Refs.~\cite{SeibergWitten}--\cite{ADNS}. Indeed, it has been
claimed~\cite{Singh2005} that as the standard linear quantum
mechanics should be a limiting case of an underlying non-linear
quantum theory, a possible approach for such a formalism can be
sought through the NC geometry. In this regard, due to changes
raised in this picture, the quantum mechanics and quantum field
theory ({\bf QFT}) phenomena have been affected and some
researches have been devoted to this
sector~\cite{Gamboa}--\cite{bertolami}. Actually, while
considering the NC geometry, the QFT modifies the relativistic
wave equation, i.e. the Dirac equation, by which, one can
calculate the consequences of NC parameters on different physical
phenomena. It is worthwhile to mention that applying the NC
effects into the action of a Dirac electron in an external
electromagnetic field, via the so called the Seiberg-Witten ({\bf
SW})  map transformations~\cite{SeibergWitten}, leads to a gauge
invariant Dirac equation~\cite{Adorno,bertolami}.

The effects of the NC geometry on electron dynamics and the ZBW
have been investigated in vacuum~\cite{zahirifarhoudi1}, and in
two-dimensional Dirac materials, and on the quantum Hall effect as
well, see, e.g., Refs.~\cite{Dayi}--\cite{bertolami2+1}. Besides,
in Ref.~\cite{Verma}, the effect of $\kappa$-deformation of
spacetime on ZBW has also been considered. Furthermore, in
Ref.~\cite{Eckstein}, it has been demonstrated that, for an
almost-commutative spacetime (it has been mentioned that this
situation is different from the NC geometry, see, e.g.,
Ref.~\cite{Dungen}), the upper bound on ZBW frequency cannot be
altered by the presence of an electromagnetic field. Also, the
effects of the NC geometry on electron dynamics in an external
magnetic field in $2+1$ dimensions have been studied in
Refs.~\cite{Liang,Panella}, wherein the Dirac equation used is not
gauge invariant~\cite{Adorno,bertolami}. Nevertheless, to our
knowledge, the effect of NC geometry on  ZBW has~not been
considered in a gauge invariant Dirac equation. Moreover, a two
dimensional Dirac equation does not accommodate the spin and
particle-hole degrees of freedom simultaneously in one picture.

In this work, we use the solutions of the gauge invariant Dirac
equation, in the presence of an external uniform magnetic field,
to study the effect of the NC geometry on the electron ZBW while
taking into account the spin degree of freedom. Accordingly, we
propose to employ the $3+1$ dimensional representation of the
Dirac equation. Hence in the next section, we review the geometry
of NC spacetime. The third section has been devoted to consider
the Dirac equation in the presence of an external uniform magnetic
field in the NC background. In the fourth section, we mainly
investigate the effects of NC geometry on the cyclotron and the
ZBW motion of an electron in this case. We study these effects via
deriving the average velocity of charge carriers for a general
wave packet in the presence of generic magnetic fields, and then,
for weak magnetic fields with a specific initial condition as
well. The summary comes at the end. Meanwhile through the work,
the lower-case Latin indices run from one to three and the Greek
indices run from zero to three.

\section{Noncommutative Geometry}
\indent

The NC geometry~\cite{Connes} particularly has interesting
features on quantum mechanics, and one of approaches toward
quantum gravity theory is deformation of the phase-space
structure, see, e.g., Ref.~\cite{Zachos}. In this geometry, the
fundamental commutators are generalized in the NC phase-space as
\begin{equation}\label{AA}
[x_{i}, x_{j}] = i \theta_{ij},\qquad\quad [p_{i}, p_{j}] = i
\beta_{ij}\qquad\quad {\rm and}\qquad\quad
[x_i,p_j]=i\hbar\left(\delta_{ij}+\sigma_{ij}\right),
\end{equation}
where the NC parameters $\theta_{ij}$ and $\beta_{ij}$ are real
constant antisymmetric sectors with dimensions of length squared
and momentum squared, respectively, and $\sigma_{ij}$s (that can
be written as a combination of $\theta_{ij}$ and $\beta_{ij}$) are
dimensionless symmetric constants. Using the Levi-Civita
antisymmetric tensor, one can write these parameters as
\begin{equation}\label{BB}
{\theta}_k\equiv {1\over
2}\,\varepsilon_{ijk}\,{\theta}_{ij}\qquad\quad {\rm and}
\qquad\quad {\beta}_k\equiv {1\over
2}\,\varepsilon_{ijk}\,{\beta}_{ij},
\end{equation}
where the parameters $\theta_k$ and ${\beta}_k$ are usually
supposed to be very small and being considered up to the
first-order~\cite{Santos2,farhoudi0}. Also, it is known that the
momentum noncommutativity plays the role of a vector potential for
a corresponding external NC uniform magnetic field, and indeed
manifests itself as a shift in
it~\cite{farhoudi0,bertolami,zahirifarhoudi1}. That is, the effect
of the NC parameters $\beta_k$ in the NC geometry is similar to
the presence of a uniform magnetic field in the usual space.
Whereas, the space noncommutativity plays an essential role when
the gauge invariance of the Dirac equation
matters~\cite{bertolami}, which is the case in this work.
Meanwhile, for the NC spacetime case of $\theta_{0 \mu}$, it has
been shown that the corresponding theory is~not
unitary~\cite{CSjT04}.

In classical physics, noncommutativity is described by replacing
the ordinary product with the Moyal star-product between two
arbitrary functions of the phase-space variables
as~\cite{SeibergWitten,Esposito}
\begin{equation}\label{B1}
(f*g)(\zeta) =\exp
\Bigl[\frac{i}{2}\alpha^{ab}\partial_a^{(1)}\partial_b^{(2)}
\Bigr]f(\zeta^{(1)})g(\zeta^{(2)}){\Bigr
|}_{\zeta^{(1)}=\zeta=\zeta^{(2)}},
\end{equation}
where $\zeta^a\equiv (x'^i,p'^j)$, $a, b=1, 2, \cdots, 2n$ in
$n$-dimensional space, and the symplectic structure is a real
matrix as
\begin{equation}\label{B2}
(\alpha_{ab})=\left(%
\begin{array}{cc}
\theta_{ij} & \delta_{ij}+\sigma_{ij} \\
-\delta_{ij}-\sigma_{ij}  &  \beta_{ij}  \\
\end{array}%
\right).
\end{equation}
However, to introduce such deformations via the Moyal product, one
instead can consider the non-canonical linear transformation
\begin{equation}\label{RR}
x_i=x'_i-{\theta_{ij}\over 2\hbar}p'_j\qquad\quad {\rm
and}\qquad\quad  p_i=p'_i+{\beta_{ij}\over 2\hbar}x'_j,
\end{equation}
which sometimes is called the Bopp-shift
method~\cite{CuFaZa,CSjT04,MaWaYa}, on the usual phase-space
variables, which also gives
$\sigma_{ij}=-\theta_{k(i}\beta_{j)l}\delta^{kl}/4$. Such a
transformation allows an extension of the commutative space to the
NC one. The main advantage of this approach is that the
Hamiltonian of system does~not need any modification. That is,
when one changes the canonical variables through (\ref{RR}), the
Hamiltonian in the NC case is still assumed to have the same
functional form as in the commutative one, i.e.
\begin{equation}\label{NCHamil}
H^{\rm [NC]}\equiv H(x_i, p_j)= H\left(x'_i-{\theta_{il}\over
2\hbar}p'_l,\, p'_j+{\beta_{jk}\over 2\hbar}x'_k\right).
\end{equation}
However, although this function is defined on the commutative
space, but the effects of NC parameters obviously arise through
their equations of motion, see, e.g.,
Refs.~\cite{RSFM,MF12,MF9,MF30}.

\section{ External Magnetic Field in NC Background}\label{sec3}
\indent

The Dirac Hamiltonian in $3+1$ dimensions, in the natural units $c=1=\hbar$, is
\begin{equation}
H_D= \left(-i\,\gamma_0 \gamma^j \partial_j +m \gamma_0\right)
\label{eq:5},
\end{equation}
where $m$ is the rest mass of electron. Meanwhile, the matrices
$\gamma_{\mu}$ are defined as\footnote{Also, in terms of the Dirac
(gamma) matrices, one usually defines
$\alpha_i\equiv\gamma_0\gamma_i$.}
\begin{eqnarray}\label{DD}
\gamma_i=\left(
                 \begin{array}{cc}
                   0 & \sigma_i \\
                  -\sigma_i & 0 \\
                 \end{array}
               \right)
               \qquad\quad\ \textrm{and}\qquad\quad\ \gamma_0=\left(
                 \begin{array}{cc}
                   I & 0 \\
                   0 & -I \\
                 \end{array}
               \right),
\end{eqnarray}
where $\sigma_i$ and $I$ are the $2\times 2$ Pauli matrices and
the unit matrix, respectively.

Now, consider the Dirac equation in the NC geometry of
space sector (i.e., when $\beta_{ij}=0$) with $\theta_{0\mu}=0$,
in the presence of an external uniform magnetic field.
Using the gauge invariant Dirac equation~\cite{Adorno}
\begin{equation}\label{sw2.1}
i\,\partial_{t}\Psi=\left( H_{\mathrm{D}}+
H_{\mathrm{SW}}^{[\theta]}\right)  \Psi ,
\end{equation}
where
\begin{equation}
H_{\mathrm{SW}}^{[\theta]}\equiv\frac{e\, Q}{2}\Big\{ \left(
\mathbf{E\times\mbox{\boldmath$\pi$}}\right)
\cdot{\mbox{\boldmath$\theta$} } +\left[
{\mbox{\boldmath$\theta$}}\times\left( {{\mbox{\boldmath$\alpha$}}
}\times\mathbf{B}\right)  \right]
\cdot\mathbf{\mbox{\boldmath$\pi$}}\Big\}.
\end{equation}
This equation is gauge invariant under $U_{\lambda
}\left(1\right)$ gauge transformations. Also,
$\mathbf{\mbox{\boldmath$\pi$}}=\mathbf{\mbox{\boldmath$p$}}-eQ\vec
A$, where $e$ is the positive unit of charge and $Q$ determines
its sign, and the corresponding Dirac Hamiltonian is
\begin{equation}
H_{\mathrm{D}}= \vec \alpha \cdot \mathbf{\mbox{\boldmath$\pi$}} +
m\gamma_0.
\end{equation}
We proceed with the case of a pure magnetic field, wherein
Eq.~(\ref{sw2.1}) reduces to
\begin{equation}\label{DiracEq1}
i\, \partial_t \Psi = \Big\{ H_{\mathrm{D}}+\frac{e\,
Q}{2}\left[{\mbox{\boldmath$\theta$}}\times\left(
{{\mbox{\boldmath$\alpha$}} }\times\mathbf{B}\right)  \right]
\cdot\mathbf{\mbox{\boldmath$\pi$}}\Big\}\,\Psi.
\end{equation}

As in
Refs.~\cite{ItzyksonZuber}--\cite{Bhattacharya}, we consider the
motion of an electron in a uniform classical background magnetic
field along the $z$-direction of the coordinate axis. That is, we
choose the background gauge field as
\begin{equation}
A^0_{\rm B} = A^y_{\rm B} = A^z_{\rm B} = 0\qquad\quad {\rm and}
\qquad\quad A^x_{\rm B} = -y{ B}. \label{GA}
\end{equation}
Furthermore, in the case of using the gauge invariant Dirac
equation, the second term in Eq.~(\ref{DiracEq1}) reads
\begin{equation}\label{DiracEq2}
H_{\mathrm{SW}}^{[\theta]}=\frac{e\, Q B}{2}
\left[\alpha_1(\pi_1\theta_3-
p_3\theta_1)+\alpha_2(p_2\theta_3-p_3\theta_2) \right].
\end{equation}
Therefore, although the Landau problem is known to be two
dimensional case~\cite{Panella}, it seems that it cannot be in the NC background, when starting from the gauge invariant Hamiltonian~(\ref{DiracEq1}).
Hence to simplify the problem, we choose $p_3=0$ and indeed assume
the momentum being in the plane perpendicular to $z$-direction. We
also take the external magnetic field being perpendicular to the
plane of momentum.

Taking these considerations into account, Eq.~(\ref{DiracEq1}) can
be solved using the spinor ansatz
\begin{equation}
\Psi = e^{-iEt} \left( \begin{array}{c} \zeta \\ \xi \end{array}
\right)
\end{equation}
with $\zeta$ and $\xi$ as 2-component objects, to get the coupled
differential equations
\begin{eqnarray}\label{diraceq}
(E-m)\zeta\!\!\!\! &=&\!\!\!\! \eta\, \vec \sigma \cdot
(-i\vec\nabla - eQ\vec A) \xi \,, \nonumber \label{eq1}\\* (E+m)
\xi\!\!\!\! &=&\!\!\!\! \eta\, \vec \sigma \cdot (-i\vec\nabla -
eQ\vec A) \zeta ,
\end{eqnarray}
where $\eta\equiv 1+ e Q B\theta_3/2$ in the natural
units\rlap.\footnote{When units are recovered $\eta\equiv
c\left(1+ e Q B\theta_3/(2\hbar)\right)$.}\ With the choice of vector
potential~(\ref{GA}), this equation is decoupled  to
\begin{equation}
(E^2 - m^2)\zeta = \eta^2 \left[ -\vec\nabla^2 + (eQ{ B})^2 y^2 -
eQ{ B}\left(2iy {\partial\over
\partial x} + \sigma_3\right) \right] \zeta .
\label{phieq0}
\end{equation}
Accordingly, one can take the solution
\begin{equation}\label{phiform}
\zeta = e^{i p_xx+i p_zz} f(y)
\end{equation}
with $f(y)$ as a 2-component matrix, and  $p_x$ and $p_z$ as the
eigenvalues of momentum.

 Also, without loss of
generality, we take $f(y)$ being the eigenstates of $\sigma_3$
with the eigenvalues $s=\pm 1$, namely
\begin{equation}\label{solution}
f^{[+]}(y) = \left(
 \begin{array}{c}
 F^{[+]}(y) \\ 0
 \end{array}
\right) \qquad\quad {\rm and}\qquad\quad f^{[-]} (y) = \left(
 \begin{array}{c}
 0 \\
 F^{[-]}(y)
 \end{array}
\right).
\end{equation}
Thus, Eq.~(\ref{phieq0}) leads to
\begin{equation}\label{Fseqn}
a^2 F^{[s]''} -  \eta^2 (eQ{ B}y + p_x)^2 F^{[s]}+ (E^2 - m^2 +
 \eta^2 eQ{B}s) F^{[s]} = 0,
\end{equation}
where prime denotes derivative with respect to the argument. With
the aid of a dimensionless variable
\begin{equation}
\chi \equiv \sqrt{e |Q| {B}} \left( y + {p_x \over eQ{ B}}
\right), \label{xi}
\end{equation}
Eq.~(\ref{Fseqn}) becomes a special form of the Hermite equation,
namely
\begin{equation}
\left( {d^2 \over d\chi^2} -\chi^2 + b^{[s]} \right) F^{[s]} = 0,
\label{diffeqn}
\end{equation}
where
\begin{equation}\label{impb0}
b^{[s]} \equiv {(E^2 - m^2)/ \eta^2 + eQ{ B}s \over e|Q|{ B}}\, .
\end{equation}
If $b^{[s]}$ being odd integers, then the solution to this
equation will exist. Hence, for $b^{[s]}=2k +1$, with $k=0, 1, 2,
\cdots $, it leads to the energy eigenvalues
\begin{equation}
E^2 = m^2 +  \eta^2 (2k+1)e|Q|{ B} -\eta^2  eQ{ B}s  \label{E}
\end{equation}
and the solutions $F_k^{[s]}=C_k e^{-\chi^2/2} H_{k}(\chi) $,
where $H_k$ is the Hermite polynomials and
$C_k\equiv\left[{\sqrt{e|Q|{ B}} / (\sqrt{\pi }k! \, 2^k) } \,
\right]^{1/2} $ is a normalization factor. In addition, the
functions $F_k^{[s]}$ satisfy the completeness relation
\begin{equation}\label{completeness}
\int_{-\infty}^{\infty} F_k^{[s]}(\chi) F_{k'}^{[s]}(\chi ) d\chi=
\sqrt{e|Q| { B}} \; \delta_{kk'}.
\end{equation}

By concentrating on the case of electrons with $Q=-1$, the energy
eigenvalues (in the natural units) are
\begin{equation}\label{En1}
E_n^{[e^{-}]}= \left(m^2  + 2n e \eta^{[e^{-}]2}
B\right)^{\frac{1}{2}},
\end{equation}
which are the relativistic forms of the Landau energy-levels with
$\eta^{[e^{-}]}= 1-e B\theta_3/2$.  Let us compare the obtained
spectrum with the corresponding one (if any) in the literature.
For simplicity, we perform it for the lowest Landau level of
(\ref{En1}) (i.e. for $n=1$) and up to the first order on the
space NC parameter, i.e. $E_1^{[e^{-}]}\simeq \left(m^2 + 2 e
B-2e^2B^2\theta_3\right)^{\frac{1}{2}}$. In this regard, we noticed
two papers~\cite{Liang,Panella} almost in this case, wherein
their results in this level are the same and (after changing their
notations and conventions to the ones used here) they obtained
$E_1^{[e^{-}]}\simeq \left(m^2 + 2 e
B+e^2B^2\theta_3\right)^{\frac{1}{2}}$. It should be noted that the starting point of these two references is the  non-gauge invariant Hamiltonian, and hence, the difference in the energy spectrum in this case is due to the gauge invariant action that we have employed.

Furthermore, as $\theta_3$ is very small\rlap,\footnote{For
example, the upper limit on $\theta_3$ has been
indicated~\cite{CSjT04,Carroll} to be $\theta_3\lesssim
10^{-40}\, {\rm m}^2 $, and/or a better bound $\theta_3\lesssim
10^{-30}\, {\rm m}^2 $~\cite{Rosa}.}\
 the energy eigenvalues actually are
\begin{equation}\label{En2}
E_n^{[e^{-}]}= \left[(m^2 + 2 n e B) - 2ne^2B^2
\theta_3\right]^{1/2}\simeq (m^2  + 2 n e B)^{1/2}-\frac{ne^2B^2
\theta_3}{(m^2  + 2 n e B)^{1/2}}\, .
\end{equation}
However in general, the energy levels are two-fold degenerates,
i.e. for $s=1$ with $n=k+1$, and for $s=-1$ with $n=k$.
Representing the solutions related to the $n$-th Landau-level by a
subscript $n$, one can thus write solutions (\ref{solution}) for
the positive energy for spin-up and spin-down as
\begin{equation}\label{fsolns}
f_{n}^{[+]} (\chi) = \left( \begin{array}{c} F_{n-1}(\chi) \\ 0
\end{array} \right)\qquad\quad {\rm and} \qquad\quad f_{n}^{[-]}  (\chi) = \left(
\begin{array}{c} 0 \\ F_n (\chi) \end{array} \right),
\end{equation}
where $F_{-1}(\chi) =0=H_{-1}(\chi)$, and hence $n\geq 1$.
Solutions~(\ref{fsolns}) determine the upper-component of the
spinor through Eq.~(\ref{phiform}). The lower-component, denoted
by $\xi$, can be solved by substituting solution $\zeta $ into the
second equation of Eqs.~(\ref{diraceq}). Finally, the positive
energy solutions of the corresponding Dirac
equations~(\ref{diraceq}) are
\begin{equation}\label{Usoln}
U_{E>0}^{[+]} (p_x,y,n) = \left(
 \begin{array}{c}
 F_{n-1}(\chi) \\[2ex] 0 \\[2ex]
 0\\[2ex]
 -\,\eta^{[e^{-}]}  {\strut\textstyle \sqrt{2ne{ B}} \over \strut\textstyle
 E_n^{[e^{-}]} +m} F_n (\chi)
 \end{array}
\right)\qquad {\rm and} \qquad U_{E>0}^{[-]}  (p_x,y,n) = \left(
 \begin{array}{c}
 0 \\[2ex] F_n (\chi) \\[2ex]
 -\, \eta^{[e^{-}]} {\strut\textstyle \sqrt{2ne{ B}} \over \strut\textstyle E_n^{[e^{-}]} +m}
 F_{n-1}(\chi) \\[2ex] 0
 \end{array}
\right),
\end{equation}
where $U^{[\pm ]}_{E>0}(p_x,y,n)$ are related to
solution~(\ref{phiform}) via solutions~(\ref{fsolns}) and the
corresponding one for $\xi$.

For the case of positrons, which are positively charged with
negative energy solutions of the Dirac equation, one has to put
$Q=1$. In this case, the energy of the relativistic Landau-levels
read
\begin{equation}\label{En3}
E_n^{[e^{+}]} = -\left(m^2  + 2n e \eta^{[e^{+}]2}
B\right)^{\frac{1}{2}}
\end{equation}
with $ \eta^{[e^{+}]}= 1+e B\theta_3/2$. Hence, it actually is
\begin{equation}\label{En4}
E_n^{[e^{+}]}=- \left[(m^2 + 2 n e B) +2ne^2B^2
\theta_3\right]^{1/2}\simeq - \left[(m^2  + 2 n e
B)^{1/2}+\frac{ne^2B^2 \theta_3}{(m^2  + 2 n e B)^{1/2}}\right]
\end{equation}
with two-fold degeneracy for $s=-1$ with $n=k+1$ and for $s=1$
with $n=k$. A similar procedure (as used for solving the positive
energy spinor) can be adopted to solve for the negative energy
spinor, and the solutions are
\begin{equation}\label{Vsoln}
U_{E<0}^{[+]}  (p_x,y,n) = \left(
 \begin{array}{c}
 0 \\[2ex]
 \eta^{[e^{+}]} {\strut\textstyle \sqrt{2ne{ B}} \over \strut\textstyle E_n^{[e^{+}]} +m}
 F_n (\hat{\chi})  \\[2ex]
 F_{n-1}(\hat{\chi}) \\[2ex] 0
 \end{array}
\right)\qquad {\rm and} \qquad U_{E<0}^{[-]}  (p_x,y,n)= \left(
 \begin{array}{c}
 \eta^{[e^{+}]} {\strut\textstyle \sqrt{2ne{ B}}
 \over \strut\textstyle E_n^{[e^{+}]} +m}
 F_{n-1}(\hat{\chi}) \\[2ex] 0  \\[2ex]
 0 \\[2ex] F_n (\hat{\chi})
 \end{array}
\right),
\end{equation}
where $U^{[\pm ]}_{E<0}(p_x,y,n)$ are also related to
solution~(\ref{phiform}) and the corresponding one for $\xi$, and
$\hat{\chi}$ is obtained from $\chi$ by changing the sign of the
$p_x$-term in definition~(\ref{xi}).

\section{ NC Effects on Cyclotron Motion and ZBW}\label{sec4}
\indent

In this section, we intend to investigate the effects of NC
geometry on the cyclotron and the ZBW motion of an electron in the
presence of an external uniform magnetic field. In this respect,
we deal with the most general wave packet, and calculate the
average velocity of charge carriers. Then, in two subsections, we
consider these effects first with generic magnetic fields and
subsequently with weak magnetic fields limit while using a
specific initial condition.

In this regard, the solution of Eq. (\ref{DiracEq1}) can be
written as a general wave packet expanded in terms of momentum
eigenfunctions via solutions (\ref{Usoln}) and (\ref{Vsoln})
as~\cite{huang1952}
\begin{equation}\label{FF}
\!\!\!\!\!\Psi(x,y,z,t)=\int \sum_{\rm n, s}c_n(p_x,z)\Bigl[e^{-i p_x x}a_s
U_{E>0}^{[s]} (p_x,y,n)\exp(-i\omega_n^{[e^{-}]}t) +e^{i p_x x} a'_s
U_{E<0}^{[s]} (p_x,y,n)\exp(i\omega_n^{[e^{+}]} t)\Bigr]d{ p}_x ,
\end{equation}
where $\omega_n^{{[e^{\pm}]}} =|E_n^{[e^{\pm}]}|$, and $a_s$ and
$a'_s$ are arbitrary coefficients. Also, following
Ref.~\cite{Romera1}, we choose $c_n(p_x,z)= \sqrt{1/(\pi
\sqrt{\sigma} D_{p_x}\, D_{z}
)}\exp\left[\frac{-(p_x^2-p_{0_{x}}^2) }
{2D_{p_x}^2}-\frac{z^2}{2D_{z}^2}
\right]\exp\left[\frac{-(n-n_0)^2}{2\sigma}\right]$ with $D_{p_x}$
and $D_z$ respectively indicating the widths of wave packet in the
corresponding momentum and spatial coordinates, $\sigma$ is the
standard deviation, and the fixed values $p_{0_{x}}$ and $n_0$ are
the expectations of distribution of momentum and Landau-level,
respectively. Actually, solution $\Psi$ is a linear combination of
the spin-up and spin-down amplitudes of the free particle Dirac
waves with positive/negative energies, whose momentum is
distributed around $p_{0_{x}}$.

Now, we derive the average of the velocity operator
$\langle\dot{{\bf
r}}\rangle=\int\Psi^{\ast}(x,y,z,t)(c\mbox{\boldmath$\alpha$})
\Psi(x,y,z,t)\,dx dy dz$. However, not to digress from the main
issue, we put the results in the Appendix. As it is obvious in
those averages (i.e., Eqs.~(\ref{averageV})), there are two types
of frequencies involved in these relations. To explore their
nature and find out how these are affected by the noncommutativity
of space sector, we look through the case when only the positive
energy charge carriers are involved. In this respect, after some
manipulations, the average velocity operators are
\begin{eqnarray}
\langle\dot{x}\rangle =
\sqrt{2eB}\,\eta^{[e^{-}]} \!\!\int \!\sum_{\rm n=1}\Big\{
 \!\!\!\!\!\! & - & \!\!\!\!\! V_{n,n+1}K_n^{[e^{-}]} F_n^2(\chi)\sqrt{n}\cos{[(\omega_{n+1}^{[e^{-}]}-
 \omega_n^{[e^{-}]})t]}\cr \nonumber \!\!\!\!\! &-&\!\!\!\!\! V_{n,n-1} K_{n-1}^{[e^{-}]} F^2_{n-1}(\chi)\sqrt{(n-1)}
 \cos{[(\omega_n^{[e^{-}]}- \omega_{n-1}^{[e^{-}]})t]} \nonumber \\  \nonumber
 \!\!\!\! \,& -& \!\!\!\!  i \,
  V_{n,n+1} K_n^{[e^{-}]} F_n^2(\chi)\sqrt{n}\sin{[(\omega_{n+1}^{[e^{-}]} -
  \omega_{n}^{[e^{-}]} )t]}\cr \!\!\!\!  &+& \!\!\!\!  i V_{n,n-1} K_{n-1}^{[e^{-}]} \,F_{n-1}^2(\chi)\sqrt{(n-1)}
   \sin{[(\omega_{n}^{[e^{-}]}-\omega_{n-1}^{[e^{-}]} )t]}\Big\} \,dy , \\ \nonumber
\end{eqnarray}
\begin{eqnarray}
\langle\dot{y}\rangle =\!\!\!\!\!\!\!\!  &&\!\!\!\!
\sqrt{2eB}\,\eta^{[e^{-}]}\!\!\int \!\sum_{\rm n=1}\Big\{ i\,
V_{n,n+1} K_n^{[e^{-}]} F_n^2(\chi)\sqrt{n}\cos{[(\omega_{n+1}^{[e^{-}]}   -
\omega_n^{[e^{-}]})t]}\cr \nonumber
\!\!\!\! & - &\!\!\!\!  i V_{n,n-1}  K_{n-1}^{[e^{-}]} \,F_{n-1}^2(\chi)\sqrt{(n-1)}
\cos{[(\omega_n^{[e^{-}]}- \omega_{n-1}^{[e^{-}]})t]}\\
\nonumber  \!\!\! & -& \!\!\!\!   V_{n,n+1}K_n^{[e^{-}]}
F_n^2(\chi)\sqrt{n}\sin{[(\omega_{n+1}^{[e^{-}]
}-\omega_{n}^{[e^{-}]} )t]}\cr \!\!\!\! &-&\!\!\!\!  V_{n,n-1}
K_{n-1}^{[e^{-}]} F_{n-1}^2(\chi)\sqrt{(n-1)}
\sin{[(\omega_{n}^{[e^{-}]}-\omega_{n-1}^{[e^{-}]} )t]}\Big\} \,
dy, \nonumber
\end{eqnarray}
and $\langle\dot{z}\rangle =0$ as expected. However, after making
a shift $n\rightarrow n+1$ in the second and forth terms of each
relation, the imaginary terms cancel out. Then, using the
completeness relation~(\ref{completeness}), these relations read
\begin{eqnarray}
\langle\dot{x}\rangle\!\!\! &= &\!\!\!\!
-2\sqrt{2eB}\sum_{\rm n=1}V_{n,n+1}\,\eta^{[e^{-}]} \!
 K_n^{[e^{-}]}  \sqrt{n}\cos{[(\omega_{n+1}^{[e^{-}]}  - \omega_n^{[e^{-}]})t]},
  \nonumber \\ \langle\dot{y}\rangle\!\!\!
&=&\!\!\!-2\sqrt{2eB}\sum_{\rm n=1} V_{n,n+1}\,\eta^{[e^{-}]}\!
K_n^{[e^{-}]}
\sqrt{n}\sin{[(\omega_{n+1}^{[e^{-}]}-\omega_{n}^{[e^{-}]} )t]}.
\end{eqnarray}
As it is expected, in this case, the frequencies, which govern the
oscillation, are the cyclotron (intraband) frequencies, i.e.
$\omega^{[\rm c]}_n\equiv\omega_{n+
1}^{[e^{\pm}]}-\omega_n^{[e^{\pm}]}$, whereas the ZBW (interband)
frequencies, $\omega^{[\rm zbw]}_n\equiv \omega_{n+
1}^{[e^{\pm}]}+\omega_n^{[e^{\mp}]}$, are absent. The result is
similar to the case of graphene~\cite{17}, wherein intraband
frequencies are due to the band quantization by the magnetic field
and are absent in the field-free situation. Whereas, the interband
frequencies, which appear due to the interference of the positive
and negative energy states, are the characteristic of ZBW.

Besides, in another work~\cite{zahirifarhoudi1}, we have shown
that, in the absence of magnetic field, neither the amplitude nor
the frequency of the ZBW motion is affected by the
noncommutativity. However, as it is obvious, when an external
uniform magnetic field is introduced, the amplitude and both the
cyclotron and the ZBW frequencies are altered. In this case, let
us find out estimations of the order of the corrections caused due
to the space noncommutativity on these frequencies in the
following subsections, where in the latter one, we also pay more
attention to the ZBW phenomenon itself.

\subsection{Generic Magnetic Fields }
\indent

In this short subsection, we just consider an external uniform
generic magnetic field and mainly focus only on the effects of the
NC term. For this purpose, using relations~(\ref{En2}) and
(\ref{En4}), and due to the smallness of the NC parameter, we have
\begin{equation}
\delta^{\rm [NC]}\omega_n^{[e^{\pm}]}\approx \pm \frac{2ne B
}{m}\delta \eta ,
\end{equation}
where $\delta \eta\equiv e B\theta_3/2 $ in the natural units.
Consequently, the NC effects on the cyclotron frequency is
$\delta^{\rm [NC]}\omega^{[c]}\approx \pm \frac{2 e B}{m}\delta
\eta$ (i.e., independent of the Landau-level), and on the ZBW
frequencies are $ \delta^{\rm [NC]} \omega_n^{[\rm zbw]} \approx
\pm 2(2n+1)\frac{e B}{m}\delta \eta$. For a generic magnetic field
of about $10\, {\rm Tesla}$, the corrections on all frequencies,
when units are recovered and with the upper bound $\theta_3\approx
10^{-30}\, {\rm m}^2 $, amount to the order of $ \approx 10^{-1}\,
{\rm s}^{-1}$, which is negligible since
\begin{eqnarray}
 \omega_n^{[\rm zbw]} +\delta^{\rm [NC]} \omega_n^{[\rm zbw]}\!\!\!\! &\approx&\!\!\!\! \omega_n^{[\rm zbw]}\left(1\pm
 10^{-22}\right),\cr
 \omega_n^{[\rm c]} + \delta^{\rm [NC]} \omega_n^{[\rm c]}\!\!\!\! &\approx&\!\!\!\!  \omega_n^{[\rm c]}\left(1\pm 10^{-14}\right).
\end{eqnarray}

However, for a very strong external magnetic field of order
$10^{10}\, {\rm Tesla}$, e.g., at the surface of neutron stars,
the corrections on all frequencies become of the order of
$10^{17}\, {\rm s}^{-1}$, which though is still small but
noticeable, namely
\begin{eqnarray}
 \omega_n^{[\rm zbw]} +\delta^{\rm [NC]} \omega_n^{[\rm zbw]}\!\!\!\! &\approx&\!\!\!\! \omega_n^{[\rm zbw]}\left(1\pm
 10^{-4}\right),\cr
 \omega_n^{[\rm c]} + \delta^{\rm [NC]} \omega_n^{[\rm c]}\!\!\!\! &\approx&\!\!\!\!
 \omega_n^{[\rm c]}\left(1\pm 10^{4}\right).
\end{eqnarray}
This result may perhaps lead to a possible detection of the NC
effects on the electron dynamics in this situation. In addition,
the amplitudes of the average velocity oscillation are also
affected by the noncommutativity of space sector, whose amount,
for a very strong external magnetic field of order $10^{10}\, {\rm
Tesla}$, is of the order $\delta \eta\approx 10^3\, {\rm
ms}^{-1}$.

\subsection{ Weak Magnetic Fields Limit }
\indent

As we have indicated in the previous subsection, an interesting
point is that the NC effect on the cyclotron and the ZBW
frequencies, and on the amplitude of the oscillations, is magnetic
field dependent. However, it is very small even for a strong
magnetic field. Thus, in this subsection, to simplify the
calculations and lending more attention to the ZBW phenomenon
itself, we choose to work in the weak magnetic fields
limit. Besides, in this approximation, as the NC effects are quite
negligible, we also neglect the corresponding terms in relations
(\ref{En2}) and (\ref{En4}). Accordingly, in this limit, it
remains
\begin{equation}
\omega_n^{[e^{\pm}]}\approx m+\frac{ n e B}{m}\equiv \omega_n.
\end{equation}
Therefore, the cyclotron frequency is independent of the
Landau-level, i.e. $\omega^{[c]}\approx e B/m$, as expected, and
the ZBW frequencies are
\begin{equation}
 \omega_n^{[\rm zbw]}
  \approx  2m+ (2n+1)\frac{e B}{m}.
\end{equation}
Furthermore, in the weak-field limit,
one has $K_n^{[e^{\pm}]}\approx 1/(2m)\equiv K$ and
$\eta^{[e^{\pm}]}\approx 1$. Also, we have neglected the terms
including $e B K_n^2$, in this limit.

To proceed while being more concrete, we choose a specific initial
condition at $t=0$ for solutions (\ref{Usoln}) that being a
localized spin-up electron in the $z$-direction in the form of a
Gaussian wave packet in phase space, while its center is at rest
in the origin, namely
\begin{equation} \label{V_Matrix_gj1}
 \Psi(x, y,z,0) =\int\sum_{n=1} c_n(p_x,z)
F_{n-1} (\chi) \left(
            \begin{array}{c}
           1 \\
            0 \\
            0 \\
            0\\
            \end{array}
          \right)e^{{-i {p}_x x}}dp_x.
\end{equation}
Thus, in this case, the wave packet~(\ref{FF}), which consists of
both positive and negative energies, reads
\begin{equation}\label{II}
\Psi(x,y,z,t)\simeq \int\sum_{n=1}
 c_n(p_x,z)\left( \!\!\!\! \begin{array}{c}
F_{n-1} (\chi)  e^{-i \omega_n t}\\[2ex] 0 \\[2ex]
0 \\[2ex]
\, K  {\strut\textstyle \sqrt{2ne{ B}}} F_{n}  (\chi)(e^{i \omega_n t}-e^{-i \omega_n t})
\end{array}\!\!\!\! \right)
            e^{{-i {p}_x x}}dp_x .
\end{equation}
This rough solution satisfies the Dirac equation when the
aforementioned approximations are taken into account. Now, let us
calculate the average velocity of charge carriers, and get
\begin{eqnarray}\label{spinupx}
\langle\dot{x}\rangle \!\!\! &\simeq &\!\!\! 2\sqrt{2eB}K
\sum_{n=1} V_{n,n+1}\sqrt{n} \Big[
 \cos{(\omega_{n+1}+\omega_{n})t}
-\cos{(\omega_{n+1}-\omega_{n})t}\Big]\nonumber \\
 \!\!\! &\simeq &\!\!\! 2\sqrt{2eB}K \sum_{\rm n=1}V_{n,n+1}
  \sqrt{n}\Big[
    \cos(\omega_n^{[\rm zbw]}t)- \cos(\omega^{[\rm c]}t)\Big]
\end{eqnarray}
and
\begin{equation}\label{spinupy}
\langle\dot{y}\rangle \simeq 2\sqrt{2eB} K\sum_{\rm n=1}
  V_{n,n+1}\sqrt{n} \Big[
    \sin(\omega^{[\rm zbw]}_n t)- \sin(\omega^{[\rm c]}t)\Big].
\end{equation}
If we specify the initial wave packet in the same procedure, but
for a localized spin-down electron as
\begin{equation}\label{V_Matrix_gj}
 \Psi(x, y,z,0) = \int\sum_{n=1} c_n(p_x,z)
F_{n} (\chi) \left(
            \begin{array}{c}
           0 \\
            1 \\
            0 \\
            0\\
            \end{array}
          \right)e^{{-i {p}_x x}}dp_x,
\end{equation}
then, we will obtain
\begin{eqnarray}\label{spindown}
\nonumber \langle\dot{x}\rangle \!\!\!&=&\!\!\!
 2\sqrt{2eB} K \sum_{\rm n=1}  V_{n,n-1}
   \sqrt{n}\Big[
    \cos(\omega^{[\rm zbw]}_nt) - \cos(\omega^{[\rm c]} t)\Big], \\
\langle\dot{y}\rangle \!\!\!&=&\!\!\! - 2\sqrt{2eB} K \sum_{\rm n=1}
   V_{n,n-1}\sqrt{n} \Big[\sin(\omega^{[\rm zbw]}_n t)+ \sin(\omega^{[\rm c]} t)\Big].
\end{eqnarray}
\begin{figure}[ht!]
\begin{minipage}[b]{0.3\linewidth}
\vspace*{0.65cm} \centering
\includegraphics[width=4cm, height=4cm]{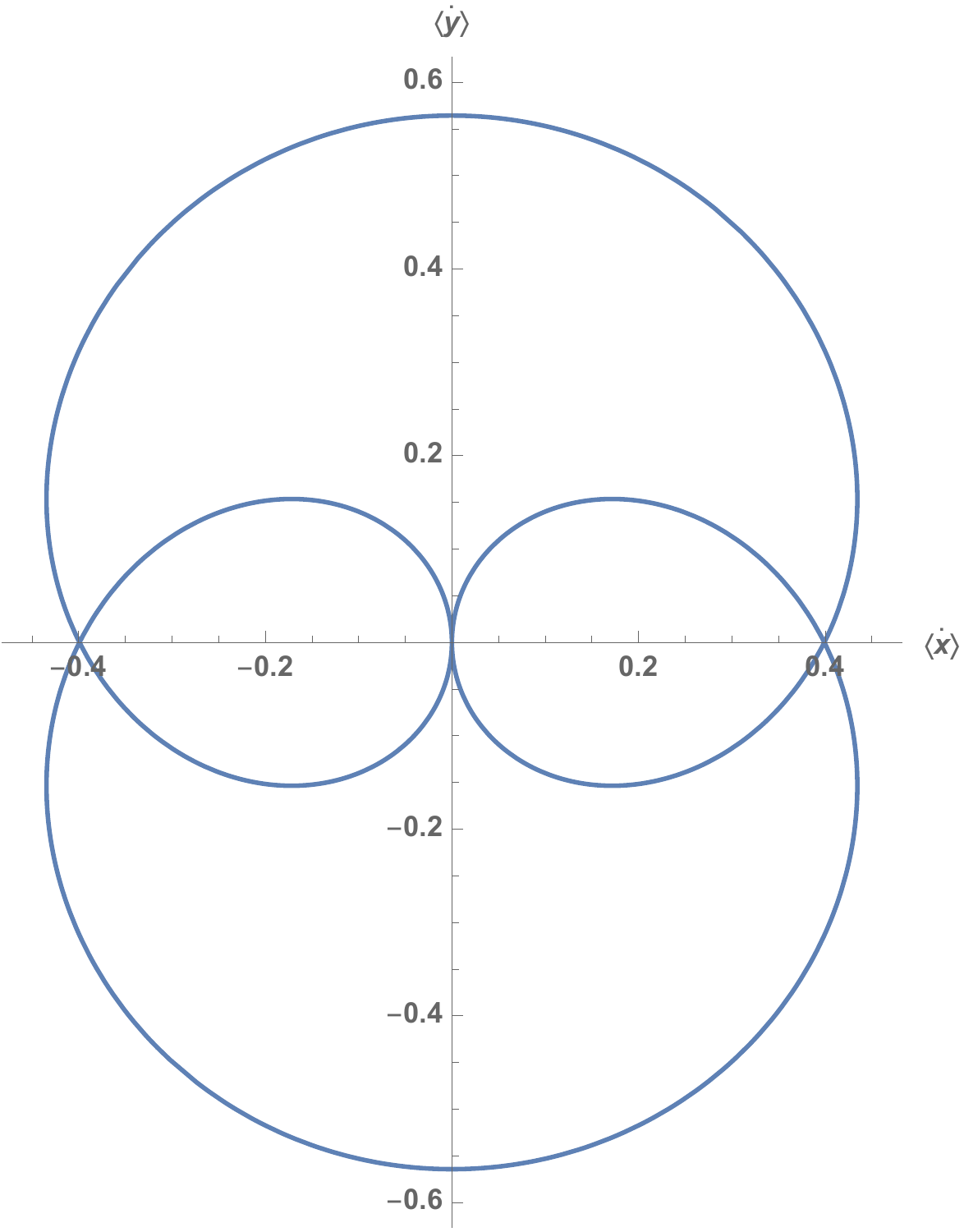}
\end{minipage}
\centering \caption{\small The trajectory of the average velocity
of charge carriers, relations~(\ref{spinupx}) and ~(\ref{spinupy})
for the initially localized spin-up electron, has been depicted
only for Landau-level centered around $n=1$ with $\sigma=3$ and
$2\sqrt{2eB} K\equiv 1$. It consists of the cyclotron and the ZBW
motions for the first $10^{-6}$ seconds.}
    \label{figef1}
\end{figure}
\begin{figure}[!]
\begin{minipage}[b]{0.3\linewidth}
\vspace*{0.65cm} \centering
\includegraphics[width=4cm, height=4cm]{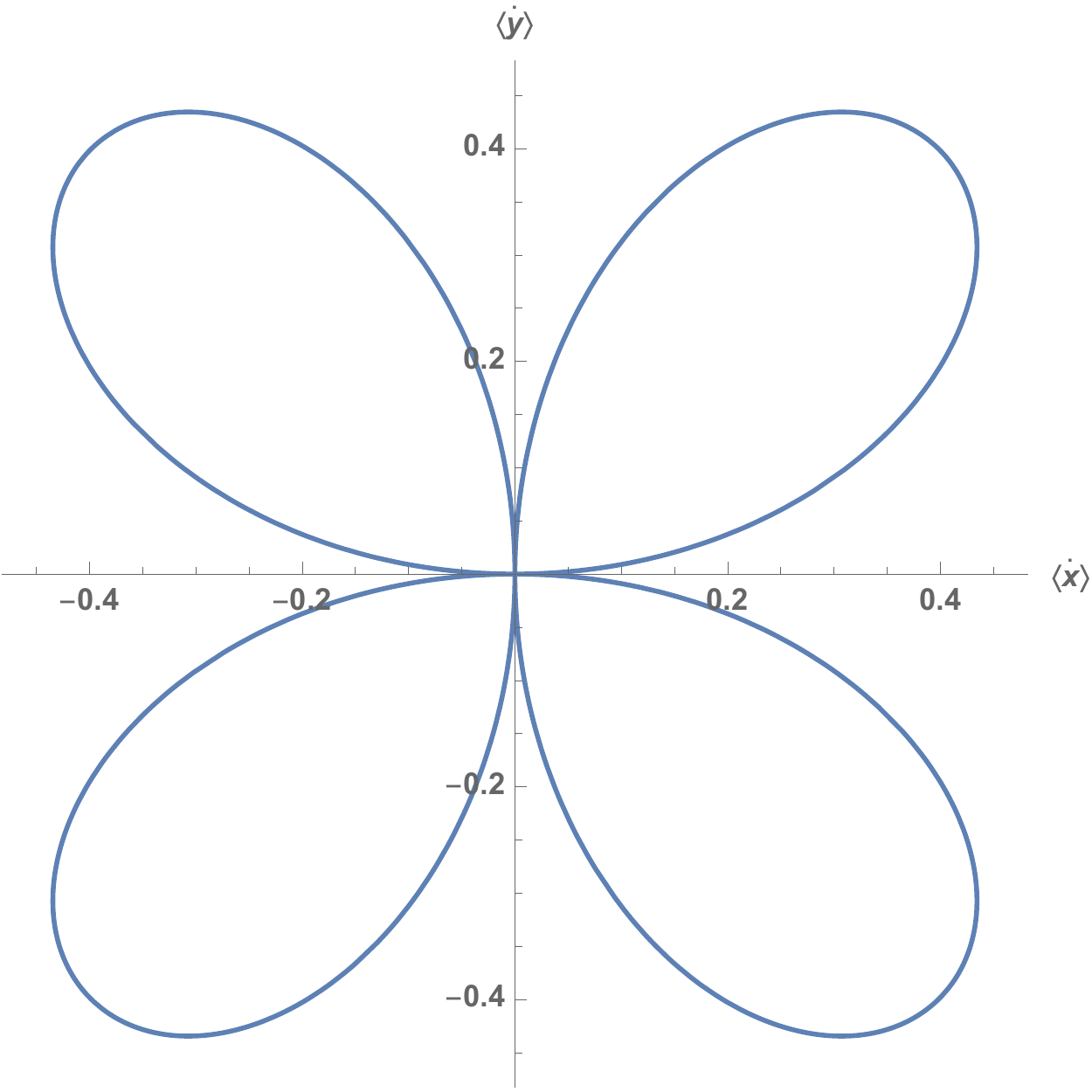}
\end{minipage}
\centering \caption{\small The trajectory of the average velocity
of charge carriers, relations~(\ref{spindown}) for the initially
localized spin-down electron, has been depicted only for
Landau-level centered around $n=1$ with $\sigma=3$ and
$2\sqrt{2eB} K\equiv 1$. It consists of the cyclotron and the ZBW
motions for the first $10^{-6}$ seconds.}
    \label{figef2}
\end{figure}

As it is obvious, in the both cases, both of the cyclotron and the
ZBW frequencies are involved. The trajectories of the average
velocity of charge carriers have been plotted in
Figs.~\ref{figef1} and~\ref{figef2} for the both initially
localized spin-up and spin-down cases. Of course, the plots have
been drawn only for Landau-level centered around $n=1$ as the main
part of the issue, with the data of $\sigma=3$ and $2\sqrt{2eB}
K\equiv 1$. The figures reveal difference between the spin-up and
spin-down cases. Besides, the ZBW phenomenon leads to a circular
motion whose direction is opposite for these two cases. Also, due
to the approximations used, the motions are unaffected from the NC
parameter. Interestingly, when electrons are prepared in the form
of wave packets, with $B\neq 0$, all the ZBW frequencies for the
Landau-levels appear. Also, both of the interband and intraband
frequencies are at work. Furthermore, although in the case at
hand, the amplitude of the oscillation is proportional to the
external magnetic field, but it is already
known~\cite{Bhattacharya} that one cannot simply put $B=0$ and
expect to get back to the free Dirac solutions.

\section{Conclusions}
\indent

The issue of dynamics of an electron in the presence of an
external uniform magnetic field in the NC background, in vacuum,
has been considered. First, the related $(3+1)$-dimensional Dirac equation, achieved from the gauge invariant action, has been solved. Then, it has been shown that the solutions
and the energy spectrum are both affected by the NC effects of the space
sector. However, the change in the energy spectrum is affected is different from previous results in Refs.~\cite{Liang,Panella}, where the non-gauge invariant  Hamiltonian has been employed.  In this regard, the ZBW motion in the NC
background, without any external magnetic field, has been
investigated~\cite{zahirifarhoudi1} before and it has been
indicated that neither the frequency nor the amplitude of the
oscillation gets affected by the NC effects. Also in
Ref.~\cite{Liang}, the change of the spectrum due to the NC
effects has been mentioned however, the solutions of the Dirac
equation and dynamics of the electron have~not been considered in
detail.

In this research, by using the extracted solutions of the Dirac
equation, we have calculated the average velocity of charge
carriers first for a general wave packet consisted of both
positive and negative energies, and then, just for a positive
energy electron. An interesting result of these parts is that both
the amplitude and the frequency of the ZBW motion are affected by
the NC space parameter. Even for a positive energy electron,
regardless of the ZBW phenomenon, the space NC effects are present
and manifest themselves as a shift in the cyclotronic motion of
electron, which, although very small, are in principle detectable.
By the way, as mentioned, since the momentum noncommutativity
leads to a NC uniform magnetic field, we have~not considered its
effects in this work.

We have given some estimations for these kinds of shifts as well.
Indeed, the calculations suggest that the stronger the external
magnetic field is (e.g., at the surface of neutron stars) the
relatively larger are the shifts of the corresponding frequencies.
In this regard, the amplitude of the oscillation also changes.
These results may pave the way for a new class of experiments that
can probably exhibit the NC effects on the motion of an electron
in a very strong external uniform magnetic field. It should be
emphasized that these effects are absent in a field-free region.

To concentrate on the ZBW phenomenon, we have also calculated the
average velocity of charge carriers in weak magnetic fields limit
for two initially localized spin-up and spin-down cases. In these
cases, due to the mentioned approximations, although the NC
effects have been neglected, the ZBW manifests itself as a
circular motion whose direction is spin dependent. This issue is
in line with studies which suggest that the spin of electron might
be generated by such a circular
motion~\cite{huang1952,zahirifarhoudi,zahirifarhoudi1} in vacuum.
We have also drawn the plots of the trajectory of the average
velocity of the charge carriers for the corresponding
Landau-levels centered around a specific level for the both cases.
The figures reveal some difference between the spin-up and
spin-down cases. Besides, the cyclotron frequency came out to have
the same direction for the both cases.

\section*{Appendix}
\indent

The averages of the components of the velocity operator are
\begin{eqnarray}\label{averageV}
\langle\dot{x}\rangle\!\!\!&=&\nonumber\\ \nonumber  \!\!\!\!\! &
& \!\!\!\!\!\! \int \sum_{\rm n,n'=1}\!\! V_{n,n'} \!\Big\{
\Big[F_{n'-1}(\hat{\chi}) F_{n}(\chi) + F_{n-1}(\hat{\chi})
F_{n'}(\chi) \Big]\cos[
    (\omega_n^{[e^{+}]} + \omega_{n'}^{[e^{-}]})t]\cr &+&\!\!\!\!\!\!
  \Big[F_{n'-1}(\chi) F_{n}(\hat{\chi}) + F_{n-1}(\chi) F_{n'}(\hat{\chi})\Big ]\cos[
    (\omega_n^{[e^{-}]} + \omega_{n'}^{[e^{+}]})t]\cr &-&\!\!\!\!\! 2 e
B \eta^{[e^{-}]}\eta^{[e^{+}]} K_n^{[e^{+}]} K_{n'}^{[e^{-}]}
\Big[ F_{n'-1}(\hat{\chi}) F_{n}(\chi) + F_{n-1}(\hat{\chi})
F_{n'}(\chi) \Big]\cos[(\omega_n^{[e^{+}]} +
\omega_{n'}^{[e^{-}]})t]\cr &-&\!\!\!\!\! 2 e B
\eta^{[e^{-}]}\eta^{[e^{+}]} K_n^{[e^{-}]} K_{n'}^{[e^{+}]}\Big[
F_{n'-1}(\chi)F_{n}(\hat{\chi}) + F_{n-1}(\chi)
F_{n'}(\hat{\chi})\Big]\cos[
    (\omega_n^{[e^{-}]} + \omega_{n'}^{[e^{+}]})t]  \\ \nonumber &- &\!\!\!\!\!
 \sqrt{2 e B}\, \eta^{[e^{-}]} K_n^{[e^{-}]}  \Big[F_{n'-1}(\chi) F_{n}(\chi) + F_{n-1}(\chi) F_{n'}(\chi)\Big]\cos[
    (\omega_n^{[e^{-}]} - \omega_{n'}^{[e^{-}]})t]\cr &+&\!\!\!\!\! \sqrt{2 e B} \eta^{[e^{+}]}K_n^{[e^{+}]} \Big[
 F_{n-1}(\hat{\chi}) F_{n'}(\hat{\chi})+F_{n'-1}(\hat{\chi}) F_{n} (\hat{\chi})\Big]\cos[
    (\omega_n^{[e^{+}]} - \omega_{n'}^{[e^{+}]})t] \\ \nonumber
    &- &\!\!\!\!\!
 \sqrt{2 e B}\,\eta^{[e^{-}]}  K_{n'}^{[e^{-}]}  \Big[F_{n'-1}(\chi) F_{n}(\chi)+ F_{n-1}(\chi) F_{n'}(\chi)\Big]\cos[
    (\omega_n^{[e^{-}]} - \omega_{n'}^{[e^{-}]})t]
 \cr &+&\!\!\!\!\! \sqrt{2 e B}\, \eta^{[e^{+}]} K_{n'}^{[e^{+}]}\Big[
 F_{n-1}(\hat{\chi}) F_{n'}(\hat{\chi})+F_{n'-1}(\hat{\chi}) F_{n} (\hat{\chi})\Big]\cos[
    (\omega_n^{[e^{+}]} - \omega_{n'}^{[e^{+}]})t] \\ \nonumber
     &-&\!\!\!\!\!
      i\Big[F_{n'-1}(\hat{\chi}) F_{n}(\chi) - F_{n-1}(\hat{\chi}) F_{n'}(\chi)\Big]\sin[
    (\omega_n^{[e^{+}]} + \omega_{n'}^{[e^{-}]})t]\cr &+&\!\!\!\!\!
i\Big[ F_{n-1}({\chi}) F_{n'}(\hat{\chi})- F_{n'-1}(\chi) F_{n}
(\hat{\chi})\Big]\sin[
    (\omega_n^{[e^{-}]} +\omega_{n'}^{[e^{+}]})t]\cr
      &+&\!\!\!\!\! 2 i e  B \eta^{[e^{-}]}\eta^{[e^{+}]} K_{n'}^{[e^{-}]}K_n^{[e^{+}]}
     \Big[F_{n'-1}(\hat{\chi}) F_{n}(\chi) - F_{n-1}(\hat{\chi}) F_{n'}(\chi)\Big]\sin[
    (\omega_n^{[e^{+}]} + \omega_{n'}^{[e^{-}]})t]\cr &+&\!\!\!\!\!
2 i e  B \eta^{[e^{-}]}\eta^{[e^{+}]}
K_n^{[e^{-}]}K_{n'}^{[e^{+}]}\Big[F_{n-1}({\chi})
F_{n'}(\hat{\chi})- F_{n'-1}(\chi) F_{n} (\hat{\chi})\Big]\sin[
    (\omega_n^{[e^{-}]} + \omega_{n'}^{[e^{+}]})t]  \\ \nonumber &+ &\!\!\!\!\! i\,\sqrt{2 e  B}\, \eta^{[e^{-}]} K_n^{[e^{-}]}
    \Big[ F_{n'-1}(\chi) F_{n}(\chi) +F_{n-1}(\chi) F_{n'}(\chi)\Big]\sin[
    (\omega_n^{[e^{-}]} - \omega_{n'}^{[e^{-}]})t]\cr &+&\!\!\!\!\! i\sqrt{2 e  B}\, \eta^{[e^{+}]} K_n^{[e^{+}]}\Big[
    F_{n-1}(\hat{\chi}) F_{n'}(\hat{\chi})+F_{n'-1}(\hat{\chi}) F_{n} (\hat{\chi})\Big]\sin[
    (\omega_n^{[e^{+}]} - \omega_{n'}^{[e^{+}]})t]\\ \nonumber &+ &\!\!\!\!\! i\,\sqrt{2 e  B}\,  \eta^{[e^{-}]}K_{n'}^{[e^{-}]}
    \Big [F_{n'-1}(\chi) F_{n}(\chi) +F_{n-1}(\chi) F_{n'}(\chi)\Big]\sin[
    (\omega_n^{[e^{-}]}- \omega_{n'}^{[e^{-}]})t]\cr &+&\!\!\!\!\! i \sqrt{2 e  B}\, \eta^{[e^{+}]} K_{n'}^{[e^{+}]}\Big
 [F_{n-1}(\hat{\chi}) F_{n'}(\hat{\chi})+F_{n'-1}(\hat{\chi}) F_{n} (\hat{\chi})\Big]\sin[
    (\omega_n^{[e^{+}]} - \omega_{n'}^{[e^{+}]})t]\Big \} \,dy, \\
\langle\dot{y}\rangle\!\!\!&=&\nonumber\\ \nonumber \!\!\!\!\! & &
\!\!\!\!\!\!
   \int \sum_{\rm n,n'=1}\!\! V_{n,n'}
   \Big\{  -\Big[ F_{n'-1}(\hat{\chi})
   F_{n}(\chi) + F_{n-1}(\hat{\chi}) F_{n'}(\chi)\Big]\sin[
    (\omega_n^{[e^{+}]} + \omega_{n'}^{[e^{-}]})t]\cr &+&\!\!\!\!\! \Big[
 F_{n-1}({\chi}) F_{n'}(\hat{\chi})+ F_{n'-1}(\chi) F_{n} (\hat{\chi})\Big]\sin[ (\omega_n^{[e^{-}]} + \omega_{n'}^{[e^{+}]})t] \cr
      \! \!\!\! & - &\!\!\!\!\! 2 e B \eta^{[e^{+}]}\eta^{[e^{-}]} K_n^{[e^{+}]}K_{n'}^{[e^{-}]}
   \Big[ F_{n'-1}(\hat{\chi}) F_{n}(\chi) + F_{n-1}(\hat{\chi}) F_{n'}(\chi) \Big]\sin[
(\omega_n^{[e^{+}]} + \omega_{n'}^{[e^{-}]})t] \cr  \! \!\!\! & +
&\!\!\!\!\! 2 e B \eta^{[e^{+}]}\eta^{[e^{-}]}
K_n^{[e^{-}]}K_{n'}^{[e^{+}]}
 \Big[F_{n-1}({\chi}) F_{n'}(\hat{\chi})+ F_{n'-1}(\chi) F_{n} (\hat{\chi})\Big]\sin[
    (\omega_n^{[e^{-}]} + \omega_{n'}^{[e^{+}]})t]  \\ \nonumber &+ &\!\!\!\!\!
  \sqrt{2 e  B}\,\eta^{[e^{-}]}  K_n^{[e^{-}]} \Big [F_{n'-1}(\chi) F_{n}(\chi) - F_{n-1}(\chi) F_{n'}(\chi)\Big]\sin[
    (\omega_n^{[e^{-}]} - \omega_{n'}^{[e^{-}]})t]\cr  &-&\!\!\!\!\! \sqrt{2 e  B}\, \eta^{[e^{+}]} K_n^{[e^{+}]} \Big
[ F_{n-1}(\hat{\chi}) F_{n'}(\hat{\chi})+F_{n'-1}(\hat{\chi})
F_{n} (\hat{\chi})\Big]\sin[
    (\omega_n^{[e^{+}]} - \omega_{n'}^{[e^{+}]})t]  \\ \nonumber &+ &\!\!\!\!\!
  \sqrt{2 e  B}\, \eta^{[e^{-}]} K_{n'}^{[e^{-}]} \Big [F_{n'-1}(\chi) F_{n}(\chi) - F_{n-1}(\chi) F_{n'}(\chi)\Big]\sin[
    (\omega_n^{[e^{-}]} - \omega_{n'}^{[e^{-}]})t] \cr & -&\!\!\!\!\!
\sqrt{2 e  B}\, \eta^{[e^{+}]} K_{n'}^{[e^{+}]} \Big[
F_{n-1}(\hat{\chi}) F_{n'}(\hat{\chi})+F_{n'-1}(\hat{\chi}) F_{n}
(\hat{\chi})\Big]\sin[
    (\omega_n^{[e^{+}]} - \omega_{n'}^{[e^{+}]})t] \\ \nonumber &-&\!\!\!\!\!
      i\Big[F_{n'-1}(\hat{\chi}) F_{n}(\chi) - F_{n-1}(\hat{\chi}) F_{n'}(\chi)\Big]\cos[
    (\omega_n^{[e^{+}]} + \omega_{n'}^{[e^{-}]})t]\cr &+&\!\!\!\!\! i\Big[-
 F_{n-1}({\chi}) F_{n'}(\hat{\chi})+ F_{n'-1}(\chi) F_{n} (\hat{\chi})\Big]\cos[  (\omega_n^{[e^{-}]} + \omega_{n'}^{[e^{+}]})t]\cr
     &+&\!\!\!\!\! 2 i e  B  \eta^{[e^{-}]}\eta^{[e^{+}]} K_{n'}^{[e^{-}]}K_n^{[e^{+}]}
     \Big[F_{n'-1}(\hat{\chi}) F_{n}(\chi) - F_{n-1}(\hat{\chi}) F_{n'}(\chi)\Big]\cos[
    (\omega_n^{[e^{+}]} + \omega_{n'}^{[e^{-}]})t] \cr &+&\!\!\!\!\!
    2 i e  B  \eta^{[e^{-}]}\eta^{[e^{+}]} K_n^{[e^{-}]}K_{n'}^{[e^{+}]}\Big[-
 F_{n-1}({\chi}) F_{n'}(\hat{\chi})+ F_{n'-1}(\chi) F_{n} (\hat{\chi})\Big]\cos[
    (\omega_n^{[e^{-}]} + \omega_{n'}^{[e^{+}]})t]  \\ \nonumber &+ &\!\!\!\!\!
   i\,\sqrt{2 e  B}\, \eta^{[e^{-}]}K_n^{[e^{-}]}
   \Big[ F_{n'-1}(\chi) F_{n}(\chi) -F_{n-1}(\chi) F_{n'}(\chi)\Big]\cos[
    (\omega_n^{[e^{-}]} - \omega_{n'}^{[e^{-}]})t] \cr &+&\!\!\!\!\!  i\,\sqrt{2 e  B}\, \eta^{[e^{+}]}K_n^{[e^{+}]}\Big
 [F_{n-1}(\hat{\chi}) F_{n'}(\hat{\chi})-F_{n'-1}(\hat{\chi}) F_{n} (\hat{\chi}))\Big]\cos[
    (\omega_n^{[e^{+}]} - \omega_{n'}^{[e^{+}]})t] \\ \nonumber &+ &\!\!\!\!\!
   i\,\sqrt{2 e  B}\, \eta^{[e^{-}]} K_{n'}^{[e^{-}]}
   \Big[F_{n'-1}(\chi) F_{n}(\chi) -F_{n-1}(\chi) F_{n'}(\chi)\Big]\cos[
    (\omega_n^{[e^{-}]} - \omega_{n'}^{[e^{-}]})t]\cr &+&\!\!\!\!\!  i\,\sqrt{2 e  B}\,  \eta^{[e^{+}]}K_{n'}^{[e^{+}]} \Big
 [F_{n-1}(\hat{\chi}) F_{n'}(\hat{\chi})-F_{n'-1}(\hat{\chi}) F_{n} (\hat{\chi})\Big]\cos[
    (\omega_n^{[e^{+}]} - \omega_{n'}^{[e^{+}]})t] \Big \}\,dy, \\ \nonumber
\langle\dot{z}\rangle\!\!\!&=&\nonumber\\ \nonumber  \!\!\!\!\! &
& \!\!\!\!\!\! \int\sum_{\rm n,n'=1}\!\!
V_{n,n'}\Big\{\Big[F_{n'-1}(\hat{\chi}) F_{n-1}(\chi) +
F_{n-1}(\hat{\chi}) F_{n'-1}(\chi)\Big]\cos[
    (\omega_n^{[e^{+}]} + \omega_{n'}^{[e^{-}]})t]\cr & -&\!\!\!\!\!
\Big[ F_{n}({\chi}) F_{n'}(\hat{\chi})+ F_{n'}(\chi) F_{n}
(\hat{\chi})\Big]\cos[ (\omega_n^{[e^{-}]} +
\omega_{n'}^{[e^{+}]})t]  \\ \nonumber &- &\!\!\!\!\! 2 e B
\eta^{[e^{-}]}\eta^{[e^{+}]} K_{n'}^{[e^{-}]}K_n^{[e^{+}]}
   \Big[F_{n'-1}(\hat{\chi}) F_{n-1}(\chi) + F_{n-1}(\hat{\chi}) F_{n'-1}(\chi) \Big ]\cos[
    (\omega_n^{[e^{+}]} + \omega_{n'}^{[e^{-}]})t]\cr &+&\!\!\!\!\!
    2 e B \eta^{[e^{-}]}\eta^{[e^{+}]} K_{n'}^{[e^{+}]}K_n^{[e^{-}]} \Big[
 F_{n}({\chi}) F_{n'}(\hat{\chi})+ F_{n'}(\chi) F_{n} (\hat{\chi})\Big]\cos[
    (\omega_n^{[e^{-}]} + \omega_{n'}^{[e^{+}]})t]  \\ \nonumber &+ &\!\!\!\!\!
  \sqrt{2 e  B}\, \eta^{[e^{-}]}K_n^{[e^{-}]} \Big[ -F_{n'-1}(\chi) F_{n-1}(\chi)
+ F_{n-1}(\hat{\chi})
F_{n'-1}(\hat{\chi})\Big]\cos[(\omega_n^{[e^{-}]}
-\omega_{n'}^{[e^{-}]})t]\cr &+&\!\!\!\!\!  \sqrt{2 e  B}\,
\eta^{[e^{+}]}K_n^{[e^{+}]}\Big[F_{n}(\chi)
F_{n'}(\chi)-F_{n'}(\hat{\chi})
  F_{n} (\hat{\chi})\Big]\cos[(\omega_n^{[e^{+}]} -\omega_{n'}^{[e^{+}]})t]\\ \nonumber &+ &\!\!\!\!\!
  \sqrt{2 e  B}\, \eta^{[e^{-}]} K_{n'}^{[e^{-}]} \Big [-F_{n'-1}(\chi) F_{n-1}(\chi)
+ F_{n-1}(\hat{\chi})
F_{n'-1}(\hat{\chi})\Big]\cos[(\omega_n^{[e^{-}]}
-\omega_{n'}^{[e^{-}]})t]\cr &+&\!\!\!\!\! \sqrt{2 e  B}\,
\eta^{[e^{+}]} K_{n'}^{[e^{+}]} \Big [F_{n}(\chi)
F_{n'}(\chi)-F_{n'}(\hat{\chi})
  F_{n} (\hat{\chi})\Big]\cos[(\omega_n^{[e^{+}]} -\omega_{n'}^{[e^{+}]})t] \\ \nonumber
     &-& \!\!\!\!\! i\Big[ F_{n'-1}(\hat{\chi})
     F_{n-1}(\chi) - F_{n-1}(\hat{\chi}) F_{n'-1}(\chi)\Big]\sin[
    (\omega_n^{[e^{+}]} + \omega_{n'}^{[e^{-}]})t]\cr &+&\!\!\!\!\! i\Big[
 F_{n}({\chi}) F_{n'}(\hat{\chi})- F_{n'}(\chi) F_{n} (\hat{\chi})\Big]\sin[
    (\omega_n^{[e^{-}]} + \omega_{n'}^{[e^{+}]})t]
    \cr &+&\!\!\!\!\! 2 i e  B \eta^{[e^{-}]}\eta^{[e^{+}]}K_n^{[e^{+}]}K_{n'}^{[e^{-}]} \Big[ F_{n'-1}(\hat{\chi})
     F_{n-1}(\chi) - F_{n-1}(\hat{\chi}) F_{n'-1}(\chi)\Big]\sin[
(\omega_n^{[e^{+}]} + \omega_{n'}^{[e^{-}]})t]\cr &-&\!\!\!\!\! 2
i e  B \eta^{[e^{-}]}\eta^{[e^{+}]}K_n^{[e^{-}]}K_{n'}^{[e^{+}]}
\Big[
 F_{n}({\chi}) F_{n'}(\hat{\chi})- F_{n'}(\chi) F_{n} (\hat{\chi})\Big]\sin[
    (\omega_n^{[e^{-}]} + \omega_{n'}^{[e^{+}]})t]  \\ \nonumber &+ &\!\!\!\!\!
  i\,\sqrt{2 e  B}\,\eta^{[e^{-}]} K_n^{[e^{-}]} \Big[F_{n'-1}(\chi) F_{n-1}(\chi)
  +F_{n-1}(\hat{\chi}) F_{n'-1}(\hat{\chi})\Big]\sin[
    (\omega_n^{[e^{-}]} - \omega_{n'}^{[e^{-}]})t]\cr &-&\!\!\!\!\!  i\,\sqrt{2 e  B} \eta^{[e^{+}]}K_n^{[e^{+}]}\Big
 [F_{n}(\chi) F_{n'}(\chi)-F_{n'}(\hat{\chi}) F_{n} (\hat{\chi})\Big]\sin[
    (\omega_n^{[e^{+}]} - \omega_{n'}^{[e^{+}]})t]\\ &+ &\!\!\!\!\!
  i\,\sqrt{2 e  B}\,\eta^{[e^{-}]}K_{n'}^{[e^{-}]} \Big[ F_{n'-1}(\chi) F_{n-1}(\chi)
  +F_{n-1}(\hat{\chi}) F_{n'-1}(\hat{\chi})\Big]\sin[
    (\omega_n^{[e^{-}]} - \omega_{n'}^{[e^{-}]})t]\cr &-&\!\!\!\!\! i\,\sqrt{2 e  B}\,\eta^{[e^{+}]}K_{n'}^{[e^{+}]} \Big
[ F_{n}(\chi) F_{n'}(\chi)-F_{n'}(\hat{\chi}) F_{n}
(\hat{\chi})\Big]\sin[
    (\omega_n^{[e^{+}]} - \omega_{n'}^{[e^{+}]})t] \Big\} \,dy,
\end{eqnarray}
where $K_{n}^{[e^{\pm}]}\equiv 1/( E_{n}^{[e^{\pm}]}+m)\, $ and
$V_{n,n'}\equiv\int c_{n'}(p_x,z)c_n(p_x,z) dp_x
dz=\frac{1}{\sqrt{\sigma}}e^{\frac{-(n-n_0)^2}{2\sigma}}e^{\frac{-(n'-n_0)^2}{2\sigma}}$.

\section*{Acknowledgements}
\indent

We thank the Research Council of Shahid Beheshti University for
financial assistance.

%
\end{document}